\documentclass{mn2e}
\usepackage{graphics}

\begin{document}

\title[Candidate $z\ge 7$ galaxies from {\em HST}/WFC3 observations of the UDF]
{The Contribution of High Redshift Galaxies to Cosmic Reionization:
New Results from Deep WFC3 Imaging of the {\em Hubble} Ultra Deep Field}
\author[Bunker et al.\ ]{Andrew J.\ Bunker\,$^{1}$,
Stephen Wilkins\,$^{1}$, Richard S.\ Ellis\,$^{2}$,
\newauthor
Daniel Stark\,$^{3}$, Silvio Lorenzoni\,$^{1}$, Kuenley Chiu\,$^{2}$, Mark Lacy\,$^{4}$
\newauthor
Matt J.\ Jarvis\,$^{5}$ \& Samantha Hickey\,$^{5}$ \\
$^{1}$\,Department of Physics, Denys Wilkinson Building, Keble Road, Oxford OX1\,3RH, U.K.\\ {\tt email:
a.bunker1@physics.ox.ac.uk}\\
$^{2}$\,California Institute of
Technology, Mail Stop 169-327, Pasadena, CA~91109, U.S.A.\\
$^{3}$\,Institute of Astrophysics, University of Cambridge,
Madingley Road, Cambridge, CB3\,0HA, U.K.\\
$^{4}$\,NRAO,  520 Edgemont Road, Charlottesville, VA 22903, U.S.A.\\
$^{5}$\,Centre for Astrophysics, Science \& Technology Research Institute, University of Hertfordshire, Hatfield, Herts AL10\,9AB, U.K.}
\date{Submitted to MNRAS}

\maketitle

\begin{abstract}
We have searched for star-forming galaxies at {$z\approx 7-10$} by applying the Lyman-break technique
to newly-released  {$Y$-, $J$- \& $H$-band images ($1.1,1.25,1.6\,\mu$m)} from WFC3 on the {\em Hubble} Space Telescope.
By comparing these images of the {\em Hubble} Ultra Deep Field with the ACS $z'$-band ($0.85\,\mu$m) images, we identify objects with red colours, $(z'-Y)_{AB}>1.3$, consistent with the Lyman-$\alpha$ forest absorption at $z\approx 6.7-8.8$. We identify 12 of these $z'$-drops down to a limiting magnitude $Y_{AB}<28.5$ (equivalent to a star formation rate of $1.3\,M_{\odot}\,{\mathrm yr}^{-1}$ at $z=7.1$), all of which are undetected in the other ACS filters. We use the WFC3 $J$-band image to eliminate contaminant low mass Galactic stars, which typically have redder colours than $z\approx 7$ galaxies.  One of our $z'$-drops is a probably a T-dwarf star. The $z\approx 7$ $z'$-drops  appear to have much bluer spectral slopes than Lyman-break galaxies at lower redshift. Our brightest $z'$-drop is not present in the NICMOS $J$-band image of the same field taken 5 years before, and is a possible transient object. From the 10 remaining $z\approx 7$ candidates we determine a lower limit on the star formation rate density of $0.0017\,M_{\odot}\,{\mathrm yr}^{-1}\,{\mathrm Mpc}^{-3}$ for a Salpeter initial mass function, which rises to $0.0025-0.004\,M_{\odot}\,{\mathrm yr}^{-1}\,{\mathrm Mpc}^{-3}$ after correction for luminosity bias. The star formation rate density is a factor of $\approx 10$ less than that of Lyman-break galaxies at $z=3-4$, and is about half the value at $z\approx 6$. {We also present the discovery of 7 $Y$-drop objects with $(Y-J)_{AB}>1.0$ and $J_{AB}<28.5$ which are candidate star-forming galaxies at higher redshifts ($z\approx 8-9$). We find no robust $J$-drop candidates at $z\approx 10$.} While based
on a single deep field, our results suggest that this star formation rate density would produce insufficient Lyman continuum photons to reionize the Universe unless the escape fraction of these photons is extremely high ($f_{\mathrm esc}>0.5$), and the clumping factor of the Universe is low. Even then, we need to invoke a large contribution from galaxies below our detection limit (a steep faint end slope). The apparent shortfall in ionizing photons might be alleviated if stellar populations at  high redshift are low metallicity or have a top-heavy initial mass function.
\end{abstract}

\section{Introduction}
\label{sec:intro}

In the past decade, the quest to observe the most distant galaxies in the
Universe has rapidly expanded to the point where the discovery of $z\simeq
6$ star-forming galaxies has now become routine.  Deep imaging surveys
with the {\em Hubble Space Telescope (HST)} and large ground based
telescopes have revealed hundreds of galaxies at $z\simeq 6$ (Bunker et
al. 2004, Yan \& Windhorst 2004, Bouwens et al. 2006, 2007, Oesch et al. 2007, Yoshida et al.
2006, McLure et al. 2009). These searches typically rely on the Lyman break galaxy (LBG) technique
pioneered by Steidel and collaborators to identify star-forming galaxies
at $z\approx 3-4$ (Steidel et al. 1996, 1999). Narrow-band searches
 have also proved successful in isolating the Lyman-$\alpha$ emission line
 in redshift slices between $z=3$ and $z=7$ (e.g., Rhoads et al.\ 2003,
 Ota et al.\ 2008, Smith \& Jarvis 2007, Ouchi et al.\ 2008), although such searches at $z>7$ have yet to yield plausible
 candidates (e.g.\ Willis et al.\ 2008). Rapid follow-up of Gamma Ray Bursts has
 also detected objects at $z\sim 6$, and most recently one at $z=8.2$ (Tanvir et al.\ 2009, Salvaterra et al.\ 2009).
 
 Parallel to these developments in identifying high redshift objects has been
 the discovery of the onset of the Gunn-Peterson (1965) effect. This is the near-total absorption of the
 continuum flux shortward of Lyman-$\alpha$ in sources at $z> 6.3$ due to the intergalactic
 medium (IGM) having a much larger neutral fraction at high redshift. The Gunn-Peterson trough was
  first discovered in the spectra of SDSS quasars (Becker et al.\ 2001, Fan et al.\ 2001, 2006). This defines the end of the reionization epoch, when the Universe
 transitioned from a neutral IGM. Latest results from WMAP indicate the mid-point of reionization may have occurred at $z\approx 11$ (Dunkley et al.\ 2009). The source
of necessary ionizing photons remains an open question: the number density
of high redshift quasars is insufficient at $z>6$ to achieve this (Fan et al.\ 2001, Dijkstra et al.\ 2004). Star-forming galaxies at high redshift are another potential driver of reionization, but we must
 first determine their rest-frame UV luminosity density to assess whether they are plausible sources;
 the escape fraction of ionizing photons from these galaxies, along with the slope of their UV spectra, are
 other important and poorly-constrained factors in determining whether star formation is responsible for the ionization of the IGM at high redshift.

Early results on the star formation rate density at $z\approx 6$ were conflicting,
with some groups claiming little to no evolution to $z\sim 3$ (Giavalisco et al.\ 2004, Bouwens
et al.\ 2003) while other work suggested that the star formation rate density at
$z\approx 6$ was significantly lower than in the well-studied LBGs at $z=3-4$ (Stanway et al.\ 2003). The consensus which has now emerged from later studies is that the abundance of {\it
luminous} galaxies is substantially {\it less} at $z\approx 6$ than at
$z\approx 3$ (Stanway et al. 2003, Bunker et al. 2004, Bouwens et al.
2006, Yoshida et al. 2006, McLure et al. 2009).  If this trend continues
to fainter systems and higher redshifts, then it may prove challenging 
for star-forming galaxies to provide the UV flux needed to fulfill
reionisation of the intergalactic medium at (e.g.\ Bunker et al. 2004). Importantly, analysis
of the faint-end of the luminosity function at high redshift has revealed that feeble
galaxies contribute an increased fraction of the total UV luminosity 
(Bouwens et al. 2006, 2007, Oesch et al. 2008, McLure et al.
2008), suggesting that the bulk of star formation (and hence reionizing
photons) 
likely come from lower luminosity galaxies not yet adequately
probed even in deep surveys.

Extending this work to the $z \approx 7$ universe has been stunted by
small survey areas (from space) and low sensitivity (from the ground),
{limiting  current $z \approx 7$ Lyman-break samples to $\sim 10$ objects} (e.g.\ Bouwens et
al.\ 2008, Oesch et al.\ 2009,  Ouchi et al.\ 2009), none of which have
been spectroscopically confirmed\footnote{One galaxy with a spectroscopic redshift of $z=6.96$ was first identified through narrow-band imaging for Lyman-$\alpha$ (Iye et al.\ 2006), but has subsequently also been identified in a survey for LBGs, Ouchi et al.\ 2009).}.
There is evidence of old stellar populations in $z\sim 4-6$  galaxies from
measurements of the Balmer break in Spitzer/IRAC imaging 
(Eyles et al.\ 2005,2007; Stark et al.\ 2007, 2009), which must have formed at higher redshift.
However, the age-dating and mass-determination of these stellar populations has many uncertainties,
so searching directly for star formation at redshifts $z\ge 7$ is critical to measure the evolution of the star formation rate density, and address the role of galaxies in reionizing the universe.
 
 The large field of view and enhanced
sensitivity of the recently-installed Wide Field Camera 3 (WFC3) on {\em HST}
has the potential to make great progress in identifying larger samples of $z>6$ Lyman break galaxies, 
as it covers an area 6.5 times that of the previous-generation NICMOS NIC3 camera,
and has better spatial sampling, better sensitivity and a filter set better tuned
to identifying high-redshift candidates through their colours. In this paper, we present an analysis of the recently-obtained WFC3 near-infrared images of the {\em Hubble} Ultra Deep Field (UDF).
We have previously used the $i'$-band and $z'$-band ACS images to identify LBGs at $z\approx 6$ through
their large $i'-z'$ colours (the $i'$-drops, Bunker et al.\ 2004). In this paper we use this ACS $z'$-band image 
in conjunction with the new WFC3 $Y$-band to search for galaxies at $z\approx 7$, with
a spectral break between these two filters (the $z'$-drops). We also anaylse the new $J$ and $H$-band WFC3 images to eliminate low-redshift contaminants of the $z'$-drop selection through their near-infrared colours, and also to look for Lyman-break galaxies at higher redshifts (the $Y$- and $J$-drops at $z\approx 8$ \& $z\approx 10$.

The structure of the paper is as follows. In Section~\ref{sec:obs} we describe
the imaging data, the construction of our catalogues and our
$z'$-drop selection. In Section~\ref{sec:discuss} we discuss the
luminosity function of star-forming sources, likely contamination
by lower-redshift interlopers, and estimate the evolution of the UV
luminosity function at $z\approx 7$. Our conclusions are
presented in Section~\ref{sec:concs}. Throughout we adopt the
standard ``concordance'' cosmology of $\Omega_M=0.3$,
$\Omega_{\Lambda}=0.7$, and use $h_{70}=H_0/70\,{\mathrm
km\,s^{-1}\,Mpc^{-1}}$. All magnitudes are on the $AB$ system (Oke
\& Gunn 1983).

\section{{\em HST}/WFC3 Observations}
\label{sec:obs}

The near-infrared images of the {\em Hubble} Ultra Deep Field (UDF) were taken
with WFC3 on {\em HST} over the period 26 August -- 6 September 2009. The infrared
channel of WFC3 was used, which is a Teledyne $1014\times 1014$ pixel HgCdTe detector,
with a field of view of $123\times 136$ (a 10-pixel strip on the edge is not illuminated by sky and used for pedestal estimation).
The data form part of the {\em HST} Treasury programme GO-11563 (P.I.\ G.~Illingworth).
This 192-orbit programme will eventually obtain F105W ($Y$), F125W ($J$) and F160W ($H$)
imaging of the UDF and two deep parallel fields, with the UDF observations comprising half
of the total orbits. The
UDF field lies within the Chandra Deep Field South (CDF-S) with the centre of the WFC3 image
located at 
coordinates RA=$03^{h}32^{m}38\fs4$, Decl.=$-27^{\circ}47'00$
(J2000), and this field is within GOODS-South area (Giavalisco et al.\ 2004), surveyed using ACS
with the same filters as the ACS UDF ($b,v,i',z'$). The WFC3 field observed falls entirely within the $11\,{\mathrm arcmin}^{2}$ of the ACS UDF images. In this paper, we focus primarily on the $Y$-band WFC3
imaging of the UDF obtained over 26--29 August 2009 U.T. This spanned 18 orbits, split into 2-orbit `visits'
with 2 pointings within each orbit. There were small offsets ($\sim< 1$\,arcsec) between the
4 exposures taken in each visit, to allow for sub-pixel stepping and to prevent bad pixels repeating. Between visits there were larger offsets of up to 10\,arcsec. The data were taken in ``MULTIACCUM" mode using SPARSAMPLE100, which non-destructively reads the array every 100\,seconds. These repeated non-destructive reads of the infrared array allow gradient-fitting to obtain the count rate (``sampling up the ramp'') and the flagging and rejection of cosmic ray strikes. Each MULTIACCUM exposure comprised 16 reads for a total duration of 1403\,sec per exposure.

To guard against the possible contamination of the high-redshift $z'$-drop sample by lower-redshift sources, we also analyse the new WFC3 $J$-band images, as lower-redshift galaxies and Galactic stars will have different $Y-J$ colours to high-redshift galaxies even if their $z'-Y$ colours are similar. The $J$-band imaging was taken over 28--30 August 2009 U.T.\ over 16 orbits, with each orbit again comprising two 1403\,sec SPARSAMPLE100 observations and a similar dithering strategy. We also measured the $(J-H)$ colours of the $z'$-drops to determine the UV spectral slopes; the WFC3 $H$-band imaging was taken from 01--05 September 2009 U.T.\ over 28 orbits and 56 SPARSAMPLE100 exposures.

\begin{figure*}
\resizebox{0.7\textwidth}{!}{\includegraphics{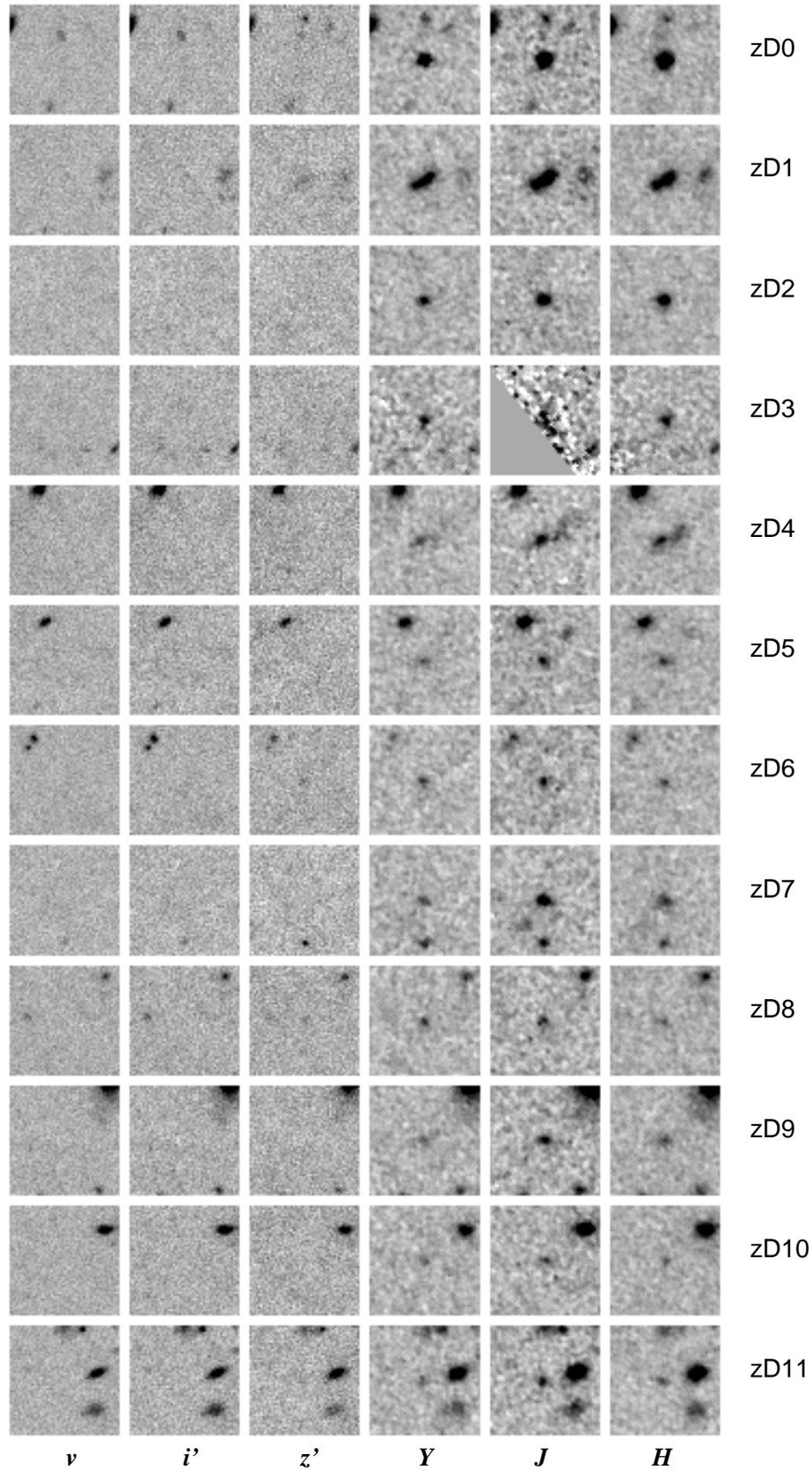}}
\caption{Postage stamp images of the $z'$-drops in the {\em Hubble} Ultra Deep Field.
The ACS $v$, $i'$, and $z'$ images are shown alongside the WFC3 $Y$-band and $J$-band 
image for each of our $z'$-drops satisfying $(z'-Y)_{AB}>1.3$, $Y_{AB}<28.5$ and with no detection in the $v$-band. Each image is 3\,arcsec across, with North up and East to the left.}
\label{fig:filters}
\end{figure*}

\begin{figure*}
\resizebox{0.68\textwidth}{!}{\includegraphics{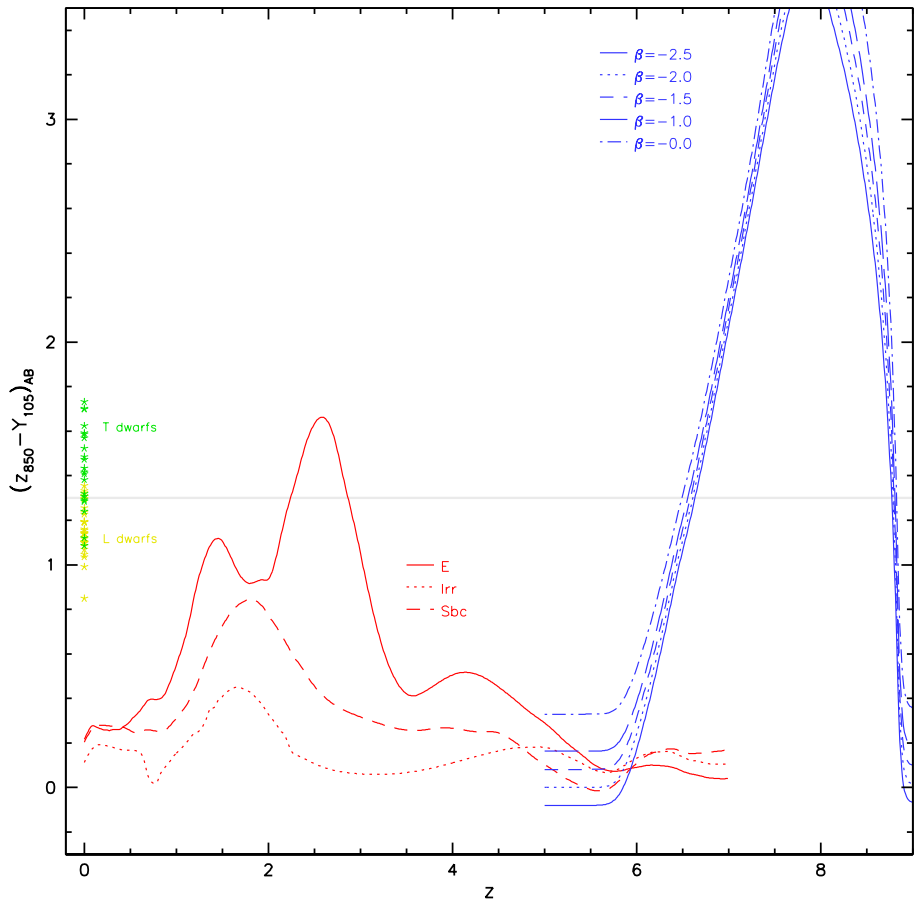}}
\caption{The redshift evolution of the $(z'-Y)$ colour of starbursting galaxies,  the extrapolated colours of low-redshift galaxies and low-mass L and T dwarf stars.  The solid, dotted, dashed, and dot-dashed blue lines show the redshift evolution of the colours of starburst galaxies with UV continuum spectral slopes $\beta\in \{-2.5,-2.0,-1.5,-1.0,0.0\}$ (i.e. $f_{\lambda}=\lambda^{\beta}$, ranging from no dust reddening to $A_{1600 {\mathrm \AA }}=4.4$ according to the empirical relation of Meurer, Heckman \& Calzetti 1999). The solid, dotted and dashed red lines show the colour of  non-evolving low-redshift galaxies (Coleman, Wu \& Weedman 1980) as a function of redshift, as three examples of foreground objects in our fields (of which elliptical galaxies at $z\approx 2.5$ would be a contaminant in the $z'$-drop selection). More discussion of contamination by Balmer/4000\,\AA\ break galaxies is given in Wilkins et al.\ (2009). The yellow and green stars denote the positions of L and T dwarf stars respectively (Chiu et al. 2006, Golimowski et al. 2004, Knapp et al. 2004).}
\label{fig:col_evol}
\end{figure*}

\begin{figure*}
\resizebox{0.82\textwidth}{!}{\includegraphics{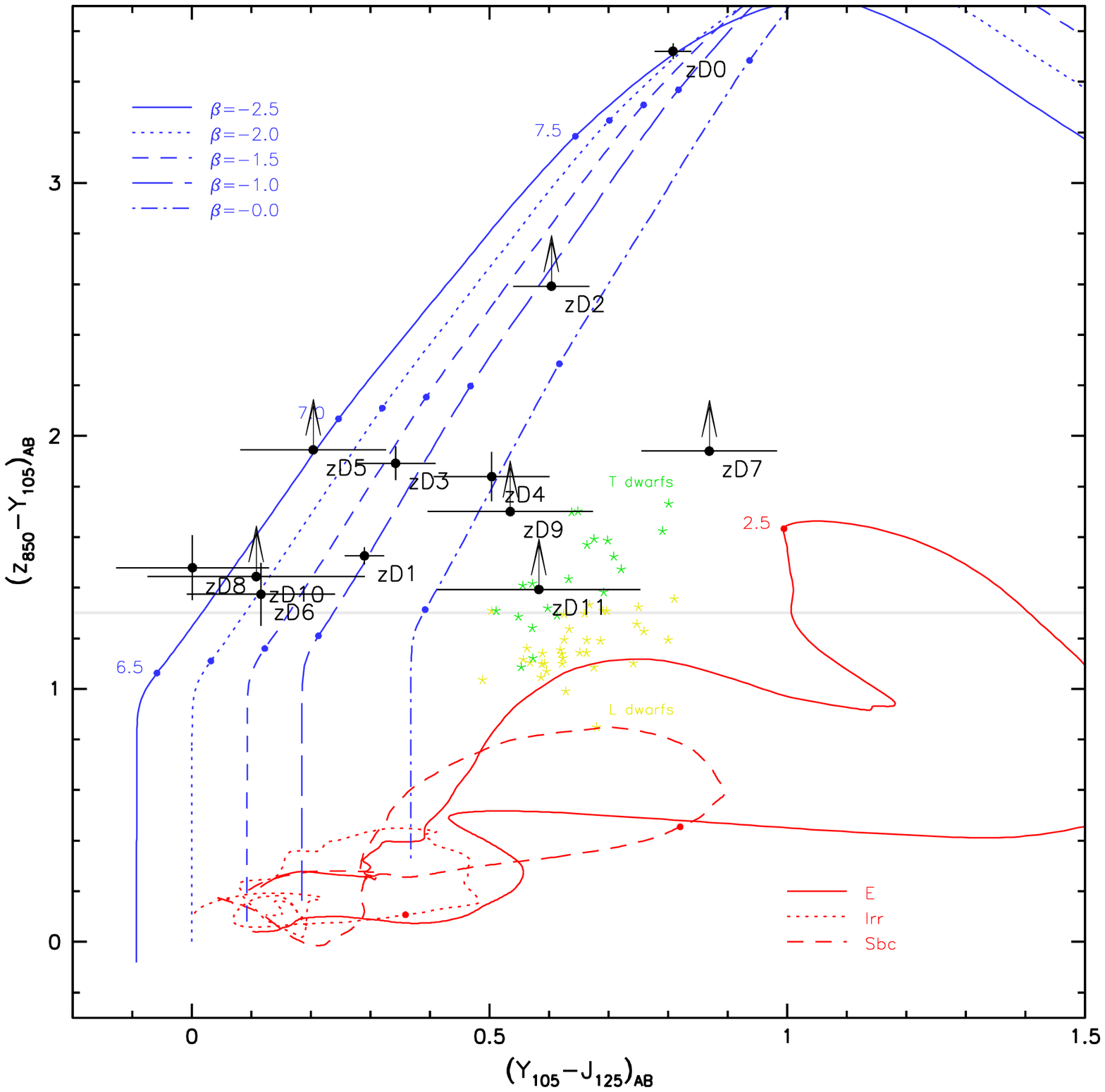}}
\caption{$(z'-Y)$ vs.\ $(Y-J)$ colour-colour diagram illustrating the differentiation of high-redshift star forming galaxies, low-mass stars and lower redshift galaxies. The black circles show the colours of $z$-dropouts in our selection and their $1\sigma$ errors (or lower limits). The solid, dotted, dashed, and dot-dashed blue lines show the evolution of the colours of starburst galaxies with UV continuum spectral slopes $\beta\in \{-2.5,-2.0,-1.5,-1.0,0.0\}$ (i.e. $f_{\lambda}=\lambda^{\beta}$ with redshift. The solid, dotted and dashed red lines show the colours of low-redshift galaxies (Coleman, Wu \& Weedman 1980) as a function of redshift. The yellow and green stars denote the positions of L and T dwarf stars respectively (Chiu et al. 2006, Golimowski et al. 2004, Knapp et al. 2004). Note that the $J$-band photometry of zD3 is suspect, as it falls close to the edge of the frame.}
\label{fig:col_stars}
\end{figure*}

\subsection{Data Reduction}
\label{sec:reduction}

The data became public on 9 September 2009, and both raw data and the output of the data reduction pipeline for individual frames made available, but not the reduced final combined images. This pipeline ({\tt stsdas.calwfc3} in IRAF) bias- and dark-subtracts the images, converts to units of electrons by multiplication by the gain, and fits the count rate to the non-destructive reads of the MULTIACCUM sequence (flagging and rejecting cosmic ray strikes in the process). The single output frame from the MULTIACCUM sequence is flatfielded, and we chose to work directly with  these pipeline outputs. Of the 9 visits in the $Y$-band observations, 2 were compromised by {image persistence --  perhaps due to observing a bright target shortly before the UDF}. These 8 frames were excluded from our combined image in case they introduced spurious sources. The remaining 28 frames, each of 1403\,sec duration, had residual background removed (which might be due to scattered light, variable dark current or small errors in the flat-fielding). This was accomplished using the {\tt XDIMSUM} package (M.~Dickinson et al.) within IRAF, which makes an average background from the disregistered (dithered) images, with bright objects masked out so as not to bias the background measurement. This average background was then subtracted from all the frames, and the shifts between the images determined from the centroids of several compact objects in the field (the header World Coordinate System was found to be insufficiently accurate to satisfactorily register the images). All the frames had the same position angle (ORIENT of 128.8 \,deg) so this shift-and-add was sufficient with no rotation required. The WFC3 point spread function at short wavelengths is undersampled by the $0\farcs13$ pixels, so we magnified each pixel onto a $3\times 3$ pixel grid so that we could shift the frames to achieve alignment at the sub-pixel level. We combined the background-subtracted, rebinned frames with {\tt imcombine} in IRAF, ignoring pixels flagged in a bad pixel mask and rejecting $4\,\sigma$ outliers. We survey 4.18\,arcmin$^2$ in all exposures (39300\,sec), and a further 0.67\,arcmin$^2$ is surveyed with about half of that exposure time. To make comparison with predicted number counts more straightforward, we restrict our analysis to the deepest 4.18\,arcmin$^2$ where the noise is approximately uniform. The $J$- and $H$-band data were reduced in an identical manner, except no frames were excluded from the combination, as the image persistence effects experienced in 22\% of the $Y$-band data were not observed in $J$- and $H$-bands. Hence the total $J$-band was 32 exposures (44900\,sec), and in $H$ it was 56 exposures (78600\,sec).

Usually undersampled imaging data from {\em HST} would be combined through geometric distortion correction and sub-pixel registration with the {\tt MULTIDRIZZLE} software (Koekemoer et al.\ 2002). However, at the time of writing, the necessary distortion files were unavailable for the new WFC3 instrument. Hence, we determined the mapping between $x,y$ pixel space in the WFC3 images, and the celestial coordinates, by determining the centroids of several hundred objects in common between the WFC3 bands and the ACS $z'$-band image of Beckwith et al.\ (2006), which had been accurately astrometrically calibrated. We use the IRAF {\tt geomap} and {\tt geotran} tasks to fit the distortions with a third order polynomial surface, producing residuals of less than $0\farcs 05$. In correcting the geometric distortions in the WFC3 images, we mapped all  the WFC3 $Y,J,H$ images and ACS $b,v,i',z'$ images to the same area, and an output pixel scale of $0\farcs06$, equivalent to $2\times 2$ binning of the original drizzled ACS pixels ($0\farcs03$).

\begin{figure}
\resizebox{0.48\textwidth}{!}{\includegraphics{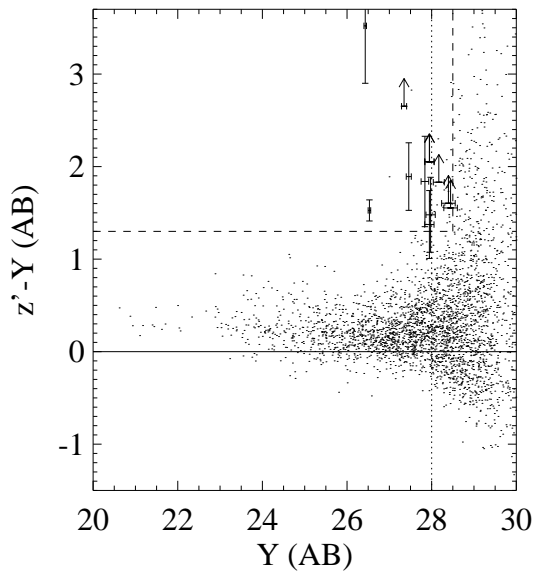}}
\caption{The $(z'-Y)_{AB}$ colours of all objects in our catalog (dots), as a function
of $Y$-band magnitude. The vertical dotted line indicates a $10\,\sigma$ detection ($Y_{AB}=28.0$).
Our primary $z'$-drop selection region where $(z'-Y)_{AB}>1.3$ and $Y_{AB}<28.5$ is marked
by the dashed lines -- candidates lie in the upper left, and those $z'$-drops which are
undetected at short wavelengths (absent from the ACS $v$-band) are marked as crosses,
or arrows denoting  $2\,\sigma$
lower limits on their colours if they are undetected in $z'$-band.}
\label{fig:colmag}
\end{figure}

The final frames had units of electrons/sec, and we take the standard ACS zeropoints for the UDF images. For the WFC3 F105W $Y$-band, {\tt synphot} reports an AB magnitude zeropoint of 26.16 (such that a source of this brightness would have a count rate of 1 electron per second). The colour:magnitude diagram (Figure~\ref{fig:colmag}) shows reasonable agreement between the $z'$-band and $Y$-band photometry, with perhaps a median colour of $0.1-0.2$\,mag (which might be due to the average galaxy colour being red, although potentially the zeropoint might also be uncertain at the $\sim$10 per cent level). The F125W $J$-band zeropoint is 26.10, and for the F160W $H$-band it is 25.81\footnote{{Note added in proof: at the time of writing (December 2009), these were still the current zeropoints in use by the MAST pipeline reduction at the Space Telescope Science Institute, using the latest calibration files. A web page associated with the WFC\,3 instrument ({\tt http://www.stsci.edu/hst/wfc3/phot\_zp\_lbn}) presents slightly different AB zeropoints of 26.27, 26.25 \& 25.96 for F105W, F125W \& F160W, which differ by 0.1-0.15\,mag from the current calibration files used in the pipeline. Adopting these does not greatly affect the selection of Lyman-break galaxies, and alters individual star formation rates by $\approx 10$ per cent, comparable to the magnitude uncertainty and less than the uncertainty on the conversion of UV flux to star formation rate.}}. We have checked the WFC3 $H$-band photometry against the NICMOS image of the same field (Thompson et al.\ 2005) taken with a very similiar F160W filter, and we see good agreement. The NICMOS wide-$J$ filter (F110W) has a significantly different transmission curve to the WFC3 F125W $J$-band, so to check the new $J$-band photometry we used instead the ESO VLT ISAAC images of GOODS-South (Vanzella et al.\ {\em in prep.}).

In our final combined $Y$-band image, we measure a FWHM of $0\farcs15$ for point sources in the field. As most high-redshift galaxies are likely to be barely resolved (e.g., Bunker et al.\ 2004, Fergusson et al.\ 2004) we perform photometry using fixed apertures of $0\farcs6$ diameter, and introduce an aperture correction to account for the flux falling outside of the aperture. This correction was determined to be 0.2\,mag in the $Y$-band  and 0.25\,mag in $J$-band and $H$-bands from photometry with larger apertures on bright but unsaturated point sources. We note that the $H$-band images display significant Airy diffraction rings around point sources. For the ACS images, the better resolution and finer pixel sampling require a smaller aperture correction of $\approx 0.1$\,mag. All the magnitudes reported in this paper have been corrected to approximate total magnitudes (valid for compact sources),and we have also corrected for the small amount of foreground Galactic extinction
toward the CDFS using the {\it COBE}/DIRBE \& {\it IRAS}/ISSA dust maps
of Schlegel, Finkbeiner \& Davis (1998). The optical reddening is
$E(B-V)=0.009$, equivalent to extinctions of
$A_{F850LP}=0.012$,  $A_{F105W}=0.010$, $A_{125W}=0.008$ \& $A_{160W}=0.005$.

The geometric transformation and image re-gridding produces an output where the noise is highly correlated. Hence measuring the standard deviation in blank areas of the final registered image will underestimate the noise, so to determine the accurate noise properties we consider
an individual frame of 1400\,sec (after flatfielding, background-subtraction and conversion
to electrons per second) where we measure a standard deviation of $\approx 0.02\,e\,{\mathrm s}^{-1}$ in
each of the combined $Y$, $J$ and $H$ images. For the final combined image, a point source of total magnitude
$Y_{AB}=28.0$ should be detected at $10\,\sigma$ in our $0\farcs6$-diameter aperture
(with the aperture correction applied). We confirmed this was the case by measuring the noise in a combined version of the data using integer pixel shifts (to avoid smoothing the data through interpolation).
The faintest sources in our catalog ($Y_{AB}=28.5$) should be detected at $6\,\sigma$.
The $J$- and $H$-band limits are nearly the same ($J_{AB}=28.0$ and $H_{AB}=27.9$ at $10\,\sigma$; and $J_{AB}=28.5$ and $H_{AB}=28.4$ at
$6\,\sigma$ for point sources measured in a $0\farcs6$-diameter aperture and with an aperture correction to total magnitude applied).
At $Y_{AB}=28.5$, a non-detection in the {\em HST} $z'$-band (which has a $2\,\sigma$ sensitivity
of $z'_{AB}=30.0$) would imply a lower limit
on the colour of $(z'-Y)_{AB}>1.5$ ($2\,\sigma$). 
We assess the recoverability and completeness as a function of magnitude by adding in fake sources to our images, and determine the fraction recovered by SExtractor with 3\,pixels (0\farcs18) of
the input source, and with a flux within 0.5\,mag. We did this by scaling down a bright but unsaturated point source and adding this in at different locations on the image -- most of the high-redshift Lyman-break galaxies show extremely compact morphologies well approximated by point sources. About 5\% of the 4.18\,arcmin$^2$ region considered was contaminated by foreground galaxies or stars and $z'$-drop candidates in these noisier regions were discounted, and our survey volume was corrected for this effect. After removing confused regions, the recoverability at our $6\,\sigma$ limit ($AB\approx 28.5$\,mag) was 92 per cent, rising to 98 per cent at our
$10\,\sigma$ limit  ($AB\approx 28.0$\,mag).
Hence the completeness corrections are negligible given our sample size and flux limit.

\subsection{Construction of Catalogs}
\label{sec:catalos}

Candidate selection for all objects in the field was performed
using version 2.5.0 of the SExtractor photometry package (Bertin
\& Arnouts 1996). For the $z'$-drops, as we are searching specifically for objects
which are only securely detected in the WFC3 $Y$-band, with minimal flux in the
ACS images, fixed circular apertures $0\farcs6$ in diameter were
`trained' in the $Y$-image and the identified apertures used to
measure the flux at the same spatial location in the $z'$-band
image by running SExtractor in dual-image mode. {For each waveband we used a weight image derived from the exposure map. This procedure} was repeated
for all other ACS and WFC3 filters. For object
identification, we adopted a limit of at least 5 contiguous pixels
above a threshold of $2\sigma$ per pixel (on the data drizzled to a
scale of 0\farcs06~pixel$^{-1}$. This cut enabled us to detect all
significant sources and a number of spurious detections close to
the noise limit, or due to diffraction spikes of stars or edge effects. Our catalog 
contains 5000 sources in all, 4000 in the region
where more than half the frames overlap; many sources are non-unique, split into multiple components
due to software deblendin). As high redshift galaxies in the rest-UV are
known to be compact (e.g., Ferguson et al.\ 2004, Bremer et al.\
2004, Bunker et al.\ 2004), we corrected the aperture magnitudes to approximate total
magnitudes with the aperture correction appropriate for that filter.

\subsection{$z\approx 7$ Candidate Selection}
\label{sec:RedshfitDiscrim}

In order to select $z\approx 7$ galaxies, we use the Lyman break technique
pioneered at $z\sim 3$ using ground-based telescopes by Steidel and
co-workers and using {\em HST} by Madau et al.\ (1996). At $z\sim 3-4$
the technique involves the use of three filters: one below the Lyman
limit ($\mathrm \lambda_{rest}=912$\,\AA ), one in the Lyman forest region
and a third longward of the Lyman-$\alpha$ line ($\mathrm
\lambda_{rest}=1216$\,\AA). We have shown that by $z\approx 6$, we can efficiently use only
two filters ($i'$- and $z'$-band for the $z\sim 6$ work), since the integrated optical depth of the Lyman-$\alpha$
forest is $\gg 1$ rendering the
shortest-wavelength filter below the Lyman limit redundant  (e.g., Bunker et al.\ 2004) . The key
issue is to work at a sufficiently-high signal-to-noise ratio that
 drop-outs can be safely identified through detection in a
single redder band. Here we extend this work to longer wavelength, and
higher redshift, by using the $Y$-band filter on WFC3 to image above
the Lyman-$\alpha$ break, and the ACS images to find the drop-out
objects. In particular, the sharp sides of $Y$-band filter (and the SDSS-type filters on ACS),
coupled the proximity in wavelength of the $Y$- and $z'$-bands,
assist in the clean selection of objects with a sharp spectral discontinuity
using this photometric redshift technique.  In
Figure~\ref{fig:col_evol} we illustrate how a
colour cut of $(z'-Y)_{AB}>1.3$ can be
effective in selecting sources with $z>6.6$. The main caveat is that at $(z'-Y)_{AB}<2.0$
we may be affected by contaminants, principally evolved galaxies at $z\approx 2.5$ (where we pick up the Balmer/4000\,\AA\ break), and low-mass cool stars in our own galaxy. These L- and T-dwarfs have colours of $0.8<(z'-Y)_{AB}<1.8$, but we may guard against these interlopers in our survey by considering the $(Y-J)$ colours which are typically much redder than $z\approx 7$ galaxies with similar $(z'-Y)_{AB}$ colours (Figure~\ref{fig:col_stars}). {We emphasize the importance in having additional ultra-deep optical  imaging from ACS ($b$, $v$ \& $i'$) in the HUDF; any object detected at these shorter wavelengths is strongly excluded from being a high-redshift source.}

Our goal is to select objects with colours redder than $(z'-Y)_{AB}>1.3$ (and preferably redder than $(z'-Y)_{AB}=1.5$, consistent with an LBG at $z \ge 7.0$). However, in order to ensure we did not exclude galaxies with colours marginally below this (but consistent with the colour cut within the magnitude uncertainties), we initially drew up a selection with a more liberal colour cut of $(z'-Y)_{AB}>1.0$. 
Imposing our selection criteria of $(z'-Y)_{AB}>1.0$ and $Y_{AB}<28.5$ on the catalog resulted in 110 objects, 24 of which were clearly spurious (mainly diffraction spikes from stars). A further 40 had intermediate colours of $1.0<(z'-Y)_{AB}<1.3$, and only 2 of these were within $1\,\sigma$ of $(z'-Y)_{AB}=1.5$ -- so this sub-sample is probably low redshift. We note that 5 objects in our intermidate colour sample, $1.0<(z'-Y)_{AB}<1.3$, are undetected in the $b,v,i'$ ACS images {(with $b_{AB}<30.3$, $v_{AB}<30.7$ and $i'_{AB}<30.6$ at $2\,\sigma$)} and may well be $i'$-drop galaxies at the high-redshift end of this selection $z\sim 6.3-6.5$, where the break occurs in the $z'$-band (leading to red $i'-z'$ and $z'-Y$ colours); indeed, two of these sources appears in the Yan \& Windhorst (2004) catalog of faint $i'$-drops in the UDF (their sources \#75 \& 85, which also appear in the catalog of Bouwens et al.\ 2006). The other 3 sources with intermediate $(z'-Y)$ colours and non-detections in the $b,v,i'$ have coordinates 03:32:41.83  $-$27:46:11.3, 03:32:37.44  $-$27:46:51.3 \& 03:32:39.58  $-$27:46:56.5, and magnitudes $Y_{AB}=28.32\pm0.09$, $27.75\pm 0.06$ \&       $27.87\pm 0.06$. We exclude these intermediate-colour objects from our $z>6.5$ selection. The remaining 46 objects with $(z'-Y)_{AB}>1.3$ were then analysed further. We demanded that these sources be undetected in the ACS optical images, as the $v$-band falls below the Lyman limit at the redshifts of interest. Twenty galaxies were detected at  $>2\,\sigma$ in $v$-band and $i'$-band (with $v_{AB}<30.7$ and $i'_{AB}<30.6$) and were eliminated as low-redshift sources. A further 3 were detected in $i'$ but not in $v$ and are probably not $z\approx 7$ galaxies. Of the remaining 23 objects, detailed examination eliminated 11 as probably spurious (due to fainter diffraction spikes, noisy areas of the image, or extended outer regions of clearly low-redshift galaxies which
had been identified by SExtractor as drop-outs due to the different PSFs between WFC3 and ACS). This left our core sample of 12 $z'$-drops. We consider whether any of these 12 might still be low redshift objects in Section~\ref{sec:Cands}.

\section{Analysis}
\label{sec:discuss}

\subsection{Discussion of Candidate $z\approx 7$ galaxies}
\label{sec:Cands}

We have 12 candidates meeting our selection criteria of $(z'-Y)_{AB}>1.3$ and 
$Y_{AB}<28.5$, after removing clearly spurious sources and those detected at shorter wavelengths in the ACS bands. We now consider whether any
of these 12 $z'$-drops are plausibly L- or T-dwarf Galactic stars. Since the resolution of WFC3 is less good than ACS, and these objects are typically undetected by  ACS, morphological cuts (i.e.\ eliminating apparently unresolved point sources) are not appropriate as they would also remove compact galaxies. Hence we use the $z'YJ$ colours to differentiate potential stars from the high-redshift galaxy tracks in colour:colour space (Figure~\ref{fig:col_stars}). Of our 12 candidates, object zD11
has colours consistent with being a T-dwarf, although we stress that it is undetected in the $z'$-band
and the $(z'-Y)_{AB}>1.5$ limit means it could still lie at high redshift. However, to be conservative, we regard this object as a probable contaminant. We note that, as our faintest $z'$-drop, if we were instead to retain it in our analysis it would have negligible impact (raising the integrated counts to $Y_{AB}=28.5$ by 10 per cent, and increasing the integrated star formation rate density by 5 per cent). All the other candidates have photometry distinct from that expected from low-mass stars, with the possible exception of zD9 which lies $>2\,\sigma$ away in colour space from the most extreme stellar colours. Both zD9 and zD11 are compact, but sufficiently faint in the near-infrared that star--galaxy separation is inaccurate; zD9 appears to have a slight East-West elongation, which would rule out the stellar interpretation.

Three of our candidates appear in previous studies (zD1, zD2 and zD3); each of these were identified previously as $z'$-band drop outs in Bouwens et al. (2004) (UDF-640-1417, UDF-983-964, UDF-387-1125), based on the shallower NICMOS imaging in F110W ($J$-wide) and F160W ($H$), along with the same UDF ACS $z'$-band as used here. The NICMOS photometry from Bouwens et al. (2004) is broadly consistent with our WFC3 measurements in $Y$ F105W, $J$ F125W and $H$, although we note the filter bandpass shape of NICMOS F110W $J$-wide is very different from the WFC3 filters. Source zD2 is also presented in Oesch et al. (2009), and source zD1 is red object \#3 in Yan \& Windhorst (2004). We have also inspected the old NICMOS images of the UDF (Thompson et al.\ 2005), and note that the majority of our candidates, except those brighter sources previously identified by Bouwens et al. (2004,2008), are undetected in the NICMOS image (which does not go as deep as the new WFC3 image). However, our brightest $z'$-drop candidate (zD0), should have easily been detected, but it is not present in either the NICMOS $J$ or $H$ images (which reach $J_{AB}=26.7$ and $H_{AB}=26.2$ at $5\,\sigma$), despite falling within the field of view and having $J_{AB}(F125W)=25.6$. This compact source is well-detected in all of the $Y$-, $J$- and $H$-band WFC3 images, so is clearly a real object rather than an artifact. It appears possible that this may be a transient object, perhaps a supernova. We note that the Yan \& Windhorst (2004) $i'$-band drop-out objects 56a\& b are the closest well-detected $z'$-band objects ($1\farcs2$ away), but it is unclear if there is a physical connection between these $z\approx 6$ sources and the transient observed with WFC3.

\subsection{UV luminosity function at $z\approx 7$}
\label{sec:lumfunc}

We use the observed surface density of our $z'$-drops in the UDF {(using a catalog cleaned
of spurious sources, and correcting the area surveyed for that masked by foreground objects, $\approx 5$ per cent)} to compare with previous estimates of the rest-UV luminosity
function at high redshift. Our final $z\sim 7$ candidate list comprised 10 objects, with
the likely star (zD11) and probable transient (zD0) removed\footnote{{Note added in proof: an independent study of the HUDF-WFC\,3 images has been posted to the preprint server (Oesch et al., submitted arXiv:0909.1806), which uses slightly bluer colour cuts and goes to lower signal-to-noise. All 10 of the $z'$-drops presented here also appear in the Oesch et al.\ sample, along with another four which are too blue for our selection (at $0.8<(z'-Y)_{AB}<1.0$, and presumably lie at slightly lower redshift than our sub-sample) and another two which are fainter than our $Y_{AB}<28.5$ cut.}}. {We used the Schechter luminosity
function with different values of faint end slope ($\alpha$), characteristic number density ($\phi^*$) and luminosity ($L^*$) to predict the observed number of sources as a function of magnitude,
which would obey our $(z'-Y)$ colour cut.} We assume a simple model spectrum for
these high-redshift star-forming galaxies, with a 99 percent flux decrement below
Lyman-$\alpha$ (i.e.\ $D_A=0.99$) due to the Lyman-$\alpha$ forest during the Gunn-Peterson era. We adopt a power law at longer wavelengths, with $f_{\lambda}\propto\lambda^{\beta}$. Measurements of the spectral slopes of $i'$-drops at $z\sim 6$ indicate rather blue colours compared to lower redshift Lyman-break galaxies (Stanway, McMahon \& Bunker 2005),
with $\beta\sim -2\rightarrow -2.2$ (i.e.\ approximately flat in $f_{\nu})$ {and indeed the $z'$-drops at $z\sim 7$ also seem to exhibit very blue spectral slopes (Figure~\ref{fig:zdrop_cols})}. Such blue colours might indicate little or no dust extinction for a constant star formation rate, according to the empiral relation $a_{1600\,{\mathrm \AA }}=4.43+1.99\beta$ (Meurer, Heckman \& Calzetti 1999), but could also be produced through a top-heavy IMF or low metallicity. For a given limiting magnitude, our code considers redshift slices over which Lyman-break galaxies would satisfy the colour cut, and computes the equivalent limiting luminosity (in terms of $L^*$) and comoving volume probed. These are then integrated over all redshifts to produce predicted number counts for comparison with our observations.

We consider three luminosity functions. The well-studied $U$-band drop-out Lyman-break galaxies at $z\sim 3$ have a rest-frame UV luminosity function with $M^*_{\mathrm 1500 \AA }=-21.1$ (equivalent to SFR$^{*}_{z=3}=15\,M_{\odot}\,{\mathrm yr}^{-1}$), $\phi^*_{\mathrm z=3}=0.00138\,{\mathrm Mpc}^{-3}$ and a faint-end slope of $\alpha=-1.6$ (Steidel et al.\ 1996, 1999). Our $z'$-drops exhibit spectral slopes close to $\beta=-2.0$ (i.e., flat in $f_{\nu}$) meaning that the $k$-correction to our observed rest-frame wavelength of $\approx 1350$\,\AA\ is negligible. We model the predicted number counts in the instance of no evolution in the properties of the star-forming population from $z\sim 7$ to $z\sim 3$. We also consider the luminosity function derived from a survey of 506 $i'$-drops at $z\sim 6$ by Bouwens et al.\ (2006), who conclude that the faint end slope is steeper ($\alpha=-1.73$), with strong evolution in the luminosity ($M^*_{z=6}=-20.25$, a factor of two fainter than $M^*_{z=3}$) and little evolution in the characteristic number density ($\phi^*_{z=6}=0.00202\,{\mathrm Mpc}^{-3}$). Finally, we consider the effects of the suggested trend of fainter $M^*$ at higher redshifts, adopting the parameterization of $M^*_{UV}=-21.0+0.36\times (z-3.8)$ suggested by Bouwens et al.\ (2009) on the basis of the paucity of $z'$-drops in  {\em HST} imaging with NICMOS (more shallow than the WFC3 data analysed here), for a predicted value of $M^*_{z=7}=-19.8$ (with $\phi^*_{z=7}=0.0011\,{\mathrm Mpc}^{-3}$ and $\alpha=-1.73$).

\begin{figure}
\resizebox{0.48\textwidth}{!}{\includegraphics{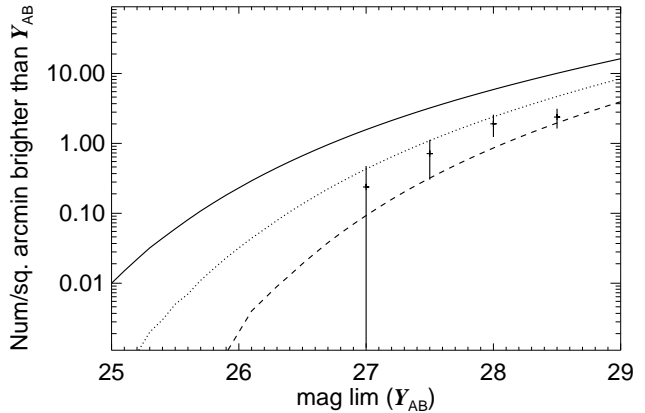}}
\caption{The observed cumulative surface density of $(z'-Y)_{AB}>1.3$ objects as a function of limiting magnitude  in $Y$-band. Overplotted are three rest-UV luminosity functions for high-redshift Lyman-break galaxy populations -- the solid curve is for the $U$-band drop-outs at $z\sim 3$ (Steidel et al.\ 1996, 1999);
and the dotted curve is for the $i'$-drops at $z\sim 6$ (Bouwens et al.\ 2006). The dashed curve represents a prediction at $z=7$ in a scenario where $M^*$ fades linearly with (1+z) (Bouwens et al.\ 2009).}
\label{fig:numdens}
\end{figure}

In Figure~\ref{fig:numdens} we compare the number counts of galaxies with colours $(z'-Y)_{AB}>1.3$ to the predictions based on the model luminosity functions at $z=3,6,7$. As can be seen, 
our observed cumulative surface density is $1.9\pm 0.7$\,arcmin$^{-2}$ brighter than $Y_{AB}=28.0$ (our 10\,$\sigma$ limit. This is well below that predicted assuming the $z=3$ UV luminosity function describes galaxies at $z\approx 7$ (it would predict 6.9\,arcmin$^{-2}$)\footnote{Recently, Reddy \& Steidel (2009) have re-fit a steeper luminosity function to the $z\approx 3$ Lyman-break population, with $\alpha=-1.73$, $\phi^*=0.00171\,{\mathrm Mpc}^{-3}$ and $M^*=-20.97$. We have recomputed the expected surface density of objects obeying our colour cuts for this luminosity function at $z\approx 7$, and the change is small: the 17\% more t $Y_{AB}<28$ and 25 per cent higher at $Y_{AB}<28.5$.}. However, our observed surface density is comparable to the luminosity function estimates assuming the $z=6$ and $z=7$ models (3.7\,arcmin$^{-2}$ and 1.0\,arcmin$^{-2}$ respectively). Pushing to fainter magnitudes (where our completeness decreases), our surface density of $z'$-drops with $Y_{AB}<28.5$ is $2.4\pm 0.7$\,arcmin$^{-2}$, which is significantly below both the $z=3$ and $z=6$ predictions of 11.5\,\&\,7.3\,arcmin$^{-2}$, and consistent with the $z=7$ model prediction of 2.2\,arcmin$^{-2}$. The volume we probe is relatively small {($\sim 10^4\,{\mathrm Mpc}^3$ comoving)}, and given the observed source counts, we estimate a cosmic variance of 35\%
assuming a pencil beam survey geometry with area 4.18 arcmin$^2$ and a
mean redshift of 7.0 with $\Delta z = 1.0$ (Trenti \& Stiavelli 2008). This uncertainty due to Cosmic Variance is comparable to the shot noise on the small number of galaxies in our sample ($1/\sqrt{10}$).

Hence we strongly rule out the simple scenario of no evolution over the range $z=7-3$ as the observed counts are 3--5 times too low, as we also established at $z\sim 6$ with the $i'$-drops (Stanway, Bunker \& McMahon 2003; Bunker et al.\ 2004). At the bright end, this interpretation agrees with the low surface density of $z'$-band drop-outs found in the HAWK-I imaging of GOODS-South presented in Hickey et al.\ (2009) of $\approx 0.01-0.02\,{\mathrm arcmin}^{-2}$ at $Y_{AB}<26.0$.We also find evidence that the $z\sim 6$ luminosity function of Bouwens et al.\ (2006) over-predicts the number of $z'$-drops at $z=7$ by a factor of $2-3$, suggesting evolution in the star forming population from $z=7-6$.

\begin{table*}
\begin{tabular}{c|c|c|c|c|clclclc}
ID & RA\,\&\,Declination J2000 &  $Y_{AB}$  & $z'_{AB}$ & $J_{AB}$ & $H_{AB}$ & $(z'-Y)_{AB}$& $(Y-J)_{AB}$    & SFR$^{\dagger}_{\mathrm UV}$ \\
\hline\hline
zD0$^{a}$ & 03:32:34.53 $-$27:47:36.0 & 26.43 $\pm$ 0.03 & 29.95 $\pm$ 0.62& 25.62 $\pm$ 0.01 & 
25.77 $\pm$ 0.02 & 3.52 $\pm$ 0.62 & 0.81 $\pm$ 0.03&  [8.82] \\ 
zD1$^{b}$ & 03:32:42.56 $-$27:46:56.6 & 26.53 $\pm$ 0.03 & 28.05 $\pm$ 0.11 & 26.23 $\pm$ 0.02& 26.26 $\pm$ 0.02 & 1.53 $\pm$ 0.11 & 0.29 $\pm$ 0.04 &  8.05 \\
zD2$^{b}$ & 03:32:38.81 $-$27:47:07.2 & 27.35 $\pm$ 0.06 & $>$30.0 (2$\sigma$)& 26.74 $\pm$ 0.03  & 26.66 $\pm$  0.03 & $>$2.65 (2$\sigma$)  & 0.60 $\pm$ 0.07 &  3.79 \\
zD3$^{b}$ & 03:32:42.57 $-$27:47:31.5 & 27.46 $\pm$ 0.06 & 29.35 $\pm$ 0.36 & 27.12$^{e}$ 
& 27.19 $\pm$ 0.05 & 1.89 $\pm$ 0.36 & 0.34 $\pm$ 0.07 & 3.42 \\
zD4 & 03:32:39.55 $-$27:47:17.5 & 27.84 $\pm$ 0.09 & 29.68 $\pm$ 0.48 & 27.34 $\pm$ 0.05 &  27.15 $\pm$ 0.05 & 1.84 $\pm$ 0.49 & 0.50 $\pm$ 0.10 &  2.41 \\
zD5 & 03:32:43.14 $-$27:46:28.5 & 27.95 $\pm$ 0.11 & $>$30.0 (2$\sigma$)   & 27.74 $\pm$ 0.07 & 27.65 $\pm$ 0.08  & $>$2.05 (2$\sigma$)  & 0.20 $\pm$ 0.13 &  2.19 \\ 
zD6 & 03:32:36.38 $-$27:47:16.2 & 27.95 $\pm$ 0.11 & 29.32 $\pm$ 0.35 & 27.83 $\pm$ 0.08 & 28.11 $\pm$ 0.12 & 1.37 $\pm$ 0.37 & 0.12 $\pm$ 0.14 &  2.18 \\
zD7 & 03:32:44.70 $-$27:46:44.3 & 27.95 $\pm$ 0.11 & $>$30.0 (2$\sigma$)   & 27.08 $\pm$ 0.04 &  27.32 $\pm$ 0.06 & $>$2.05 (2$\sigma$)  & 0.87 $\pm$ 0.12 &  2.18 \\ 
zD8 & 03:32:40.57 $-$27:46:43.6 & 27.98 $\pm$ 0.11 & 29.45 $\pm$ 0.39 & 27.97 $\pm$ 0.09 & 28.33 $\pm$ 0.15 & 1.48 $\pm$ 0.41 & 0.01 $\pm$ 0.14 &  2.13 \\ 
zD9$^{c}$  & 03:32:37.21 $-$27:48:06.2 & 28.17 $\pm$ 0.13 & $>$30.0 (2$\sigma$)   & 27.63 $\pm$ 0.07 & 27.70 $\pm$ 0.08 & $>$1.83 (2$\sigma$)  & 0.53 $\pm$ 0.15 &  1.78 \\ 
zD10 & 03:32:39.73 $-$27:46:21.3 & 28.40 $\pm$ 0.16 & $>$30.0 (2$\sigma$)   & 28.29 $\pm$ 0.12 & 28.53 $\pm$ 0.18 & $>$1.6 (2$\sigma$)  & 0.11 $\pm$ 0.20 &  1.44 \\ 
zD11$^{d}$ & 03:32:38.36 $-$27:46:11.9 & 28.45 $\pm$ 0.16 & $>$30.0 (2$\sigma$)   & 27.86 $\pm$ 0.08 & 28.33 $\pm$ 0.15 & $>$1.55 (2$\sigma$)  & 0.58 $\pm$ 0.18 &  [1.38] \\
\end{tabular}
$^{\dagger}$ The star formation rate (SFR) is in units of $M_{\odot}\,{\mathrm yr}^{-1}$ based on a relation between the rest-UV continuum density of $L_{1500\,{\mathrm \AA\ }}=8\times 10^{27} {\mathrm erg\,s}^{-1}\,/\,{\mathrm SFR}$,
where we assume that the objects are at the predicted average redshift of the $z'$-drops ($\bar{z}=7.1$). Where the SFR is in brackets, these objects are unlikely to be at $z\approx 7$, but the equivalent star formation rate for their $Y$-band magnitudes is listed but not used in our calculation of the star formation rate density.
\caption{$z$-band drop out candidates meeting our selection criteria with $Y<28.5$ AB mag. $^{a}$ likely a SN. $^{b}$ identified as $z$-band drop outs in previous studies. $^{c}$ possibly a star based on their $Y_{105}-J_{125}$ colour. $^{d}$ likely to be a star based on their $Y_{105}-J_{125}$ colour. $^{e}$ The $J$-band photometry of this source (zD3) is compromised as it falls on the edge of the frame.}
\label{tab}
\end{table*}

\begin{table*}
\begin{tabular}{c|c|c|c|c|clclc}
ID & RA\,\&\,Declination J2000 &  $J_{AB}$  & $Y_{AB}$ & $H_{AB}$ & $(Y-J)_{AB}$ & $(J-H)_{AB}$& $r_{hl}$ / arcsec \\
\hline\hline
YD1 &   03:32:42.88  $-$27:46:34.5
 &   27.70 $\pm$   0.08 &   28.89 $\pm$   0.24 &  27.98 $\pm$ 0.10 &  1.19 $\pm$   0.25 &  $-$0.28 $\pm$ 0.13 & 0.38  \\ 
YD2 &   03:32:37.80  $-$27:46:00.1
 &   27.88 $\pm$   0.10 &   30.12 $\pm$   0.73 &  28.07 $\pm$ 0.11 &  2.24 $\pm$   0.73 & $-$0.19 $\pm$ 0.14 &  0.14  \\ 
YD3 &   03:32:38.14  $-$27:45:54.0
 &   28.07 $\pm$   0.11 &   29.77 $\pm$   0.53 &  28.07 $\pm$ 0.11 &  1.70 $\pm$   0.54 & 0.00 $\pm$ 0.15 &  0.16  \\ 
YD4 &   03:32:33.13  $-$27:46:54.4
 &   28.11 $\pm$   0.11 &   29.74 $\pm$   0.66 &  28.9 $\pm$ 0.23 &  1.63 $\pm$   0.67 & $-$0.79 $\pm$ 0.26 & 0.11  \\ 
YD5 &   03:32:35.85  $-$27:47:17.1
 &   28.38 $\pm$   0.14 &   29.44 $\pm$   0.39 &  27.98 $\pm$ 0.10 &  1.06 $\pm$   0.42 & 0.41 $\pm$ 0.17 &  0.10  \\ 
YD6 &   03:32:40.40  $-$27:47:18.8
 &   28.40 $\pm$   0.16 &   29.99 $\pm$   0.67 &  28.07 $\pm$ 0.19 &  1.59 $\pm$   0.69 & $-$0.31 $\pm$ 0.25 &  0.11  \\ 
YD7 &   03:32:37.63  $-$27:46:01.5
 &   28.44 $\pm$   0.16 &   29.57 $\pm$   0.43 &  28.61 $\pm$ 017 &  1.13 $\pm$   0.46 & $-$0.17 $\pm$ 0.24 &  0.10  \\ 
\end{tabular}
\caption{Coordinates and magnitudes of the $Y$-band drop out candidates meeting our selection criteria with $(Y-J)_{AB}>1.0$, $J<28.5$ AB mag and no detection at $>2\,\sigma$ in the ACS images. We also present the measured half-light radius (for an unresolved source, $r_{hl}=0\farcs08$).}
\label{tab:ydrops}
\end{table*}

\subsection{The Star Formation Rate Density at $z\approx 7$ and Implications for Reionization}
\label{sec:SFRD}

We can use the observed $Y$-band magnitudes of objects in the $z'$-drop sample to estimate their star formation rate from the rest-frame UV luminosity density. 
In the absence of dust
obscuration, the relation between the flux density in the rest-UV
around $\approx 1500$\,\AA\ and the star formation rate (${\mathrm SFR}$
in $M_{\odot}\,{\mathrm yr}^{-1}$) is given by $L_{\mathrm UV}=8\times 10^{27}
{\mathrm SFR}\,{\mathrm ergs\,s^{-1}\,Hz^{-1}}$ from Madau, Pozzetti \&
Dickinson (1998) for a Salpeter (1955) stellar initial mass function
(IMF) with $0.1\,M_{\odot}<M^{*}<125\,M_{\odot}$. This is comparable
to the relation derived from the models of Leitherer \& Heckman (1995)
and Kennicutt (1998).  However, if a Scalo (1986) IMF is used, the
inferred star formation rates will be a factor of $\approx 2.5$ higher
for a similar mass range.
In the absence of a spectroscopic redshift, we assume that these lie at the predicted average redshift for galaxies obeying our colour cuts and with $Y_{AB}<28.5$. For the luminosity functions considered, the predicted mean redshift is around $\bar{z}=7.1$ for a spectral slope $\beta \approx -2.0$. 
At the limit of our catalog ($Y_{AB}=28.5$), the inferred star formation rate for a galaxy at $z=7.1$ is $1.3\,M_{\odot}\,{\mathrm yr}^{-1}$, equivalent to $0.1\,L^*$ for $M^*_{UV}=-21.1$ (as at $z=3$) or $0.2\,L^*$ for $M^*_{UV}=-20.24$ (for the $z=6$ luminosity function of Bouwens et al.\ 2007). The total star formation rate in our 10 candidate $z\approx 7$ galaxies is $29.6\,M_{\odot}\,{\mathrm yr}^{-1}$ assuming they are at the predicted mean redshift. The colour selection of $(z'-Y)$ is effective over the redshift range $6.7<z<8.8$ (Figure~\ref{fig:col_evol}), and so the 4.2\,arcmin$^{2}$ surveyed in the deepest region of the WFC3 $Y$-image should colour-select $z'$-drop galaxies within a comoving volume of 17700\,Mpc$^{3}$. So we calculate the star formation rate density from the $z'$-drops in our sample
to be $0.0017\,M_{\odot}\,{\mathrm yr}^{-1}\,{\mathrm Mpc}^{-3}$. However, this should be regarded as a conservative {\em lower limit} on the star formation rate density, because the redshift range is not surveyed with uniform sensitivity to UV luminosity (and hence star formation rate). We have a strong luminosity bias towards the lower end of the redshift range, due to the effects of increasing luminosity distance with redshift and also the Lyman-$\alpha$ break obscuring an increasing fraction of the filter bandwidth. Using an ``effective volume" approach (e.g., Steidel et al.\ 1999) accounts for this luminosity bias, and increases the inferred total star formation rate density to $0.0034\,M_{\odot}\,{\mathrm yr}^{-1}\,{\mathrm Mpc}^{-3}$, integrating down to $0.2\,L^*_{UV}$ of the $L^*$ value at $z=3$ (i.e.\ $M_{UV}=-19.35$)  and assuming a UV spectral slope of $\beta=-2.0$.  If instead we integrate down to $0.1\,L^*_{z=3}$ ($0.2\,L^*_{z=6}$) and the total star formation rate density is $0.004\,M_{\odot}\,{\mathrm yr}^{-1}\,{\mathrm Mpc}^{-3}$. These star formation rate densities are a factor of $\sim 10$ {\em lower} than at $z\sim 3-4$, and even a factor of $1.5-3$ below that at $z\approx 6$ (Bunker et al.\ 2004; Bouwens et al.\ 2006).

 We can compare our measured UV luminosity density at $z\approx 7$ (quoted above as a corresponding star formation rate) with that required to ionize the Universe at this redshift.
 Madau, Haardt \& Rees (1999) give the density of star formation
required for reionization (assuming the same Salpeter IMF as used in this paper):
\[
 {\dot{\rho}}_{\mathrm SFR}\approx \frac{0.005\,M_{\odot}\,{\mathrm yr}^{-1}\,{\mathrm Mpc}^{
-3}}{f_{\mathrm esc}}\,\left( \frac{1+z}{8}\right) ^{3}\,\left( \frac{\Omega_{b}\,h^
2_{70}}{0.0457}\right) ^{2}\,\left( \frac{C}{5}\right)
 \]
We have updated equation 27 of Madau, Haardt \& Rees (1999) for a more recent concordance cosmology estimate of the baryon
density from Spergel et al.\ (2003).

\begin{figure}
 \resizebox{0.5\textwidth}{!}{\includegraphics{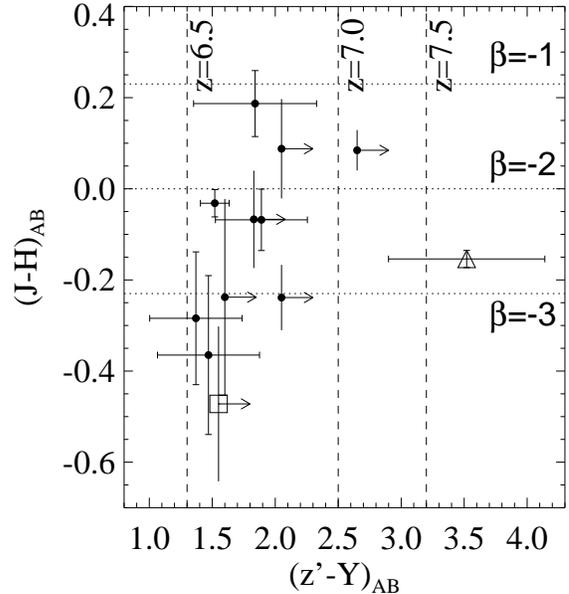}}
\caption{The near-infrared colours of the $z'$-drops in the {\em Hubble} Ultra Deep Field.
Many have very blue $(J-H)$ colours. The dotted horizontal lines show the $(J-H)$ colours
produced by spectral slopes of $\beta =-1,-2,-3$, and the dashed vertical lines show the redshifts
which would produce the $(z'-Y)$ colours, assuming no emission line contamination from Lyman-$\alpha$ and a spectral slope of $\beta=-2.0$. The triangle is the probably transient source (zD0) and the square is the likely T-dwarf star (zD11).}
\label{fig:zdrop_cols}
\end{figure}

$C$ is the clumping factor of neutral
hydrogen, $C=\left< \rho^{2}_{\mathrm HI}\right> \left< \rho_{\mathrm
    HI}\right> ^{-2}$. Early simulations suggested $C\approx 30$ (Gnedin \&
Ostriker 1997), but more recent work including the effects of reheating implies a lower
concentration factor of  $C\approx 5$ (Pawlik et al.\ 2009). 
The escape fraction of ionizing photons ($f_{\mathrm esc}$) for
high-redshift galaxies is highly uncertain (e.g., Steidel, Pettini \&
Adelberger 2001, Shapley et al.\ 2006), but even if we take $f_{\mathrm esc}=1$ (no absorption by
H{\scriptsize~I}) and a very low clumping factor, this estimate of the star formation density required
is a factor of 1.5--2 higher than our measured star formation density at
$z\approx 7$ from $z'$-drop galaxies in the UDF.  For faint end slopes of $\alpha ~-1.8\rightarrow-1.6$
galaxies with $L>0.2\,L^{*}$ account for $24-44$\% of the total
luminosity (if there is no low-luminosity cut-off for the Schechter function), so even with a steep faint-end slope  at $z\approx 7$ we still fall short of the required
density of Lyman continuum photons required to reionize the Universe, unless the escape fraction is implausibly high ($f_{\mathrm esc}>0.5$) and/or the faint end slope is $\alpha<-1.9$ (much steeper than observed at $z=0-6$).
However, the assumption of a solar metallicity Salpeter IMF may be flawed: the colours of $z\sim 6$ $i'$-band drop-outs are very blue (Stanway, McMahon \& Bunker 2005), with $\beta<-2$, and the new WFC3 $J$- and $H$-band images show that the $z\approx 7$ $z'$-drops also have blue colours on average (Figure~\ref{fig:zdrop_cols}). A slope of $\beta<-2$ is bluer that for continuous star formation with a Salpeter IMF, even if there is no dust reddening. Such blue slopes could be explained through low metallicity,
 or a top-heavy IMF, which can produce between 3 and 10 times as many ionizing photons for the same 
 1500\,\AA\ UV luminosity (Schaerer 2003 -- see also Stiavelli, Fall \& Panagia 2004). Alternatively, we may be seeing galaxies at the onset of star formation, or with a rising star formation rate (Verma et al.\ 2007), which would also lead us to underestimate the true star formation rate from the rest-UV luminosity. We explore the implications of the blue UV spectral slopes in $z\ge 6$ galaxies in a forthcoming paper (Wilkins et al.\ {\em in prep.}).
  
  \subsection{Star-Forming Galaxies at $z\approx 8$ and Beyond}
\label{sec:Jdrops}

\begin{figure*}
\resizebox{0.95\textwidth}{!}{\includegraphics{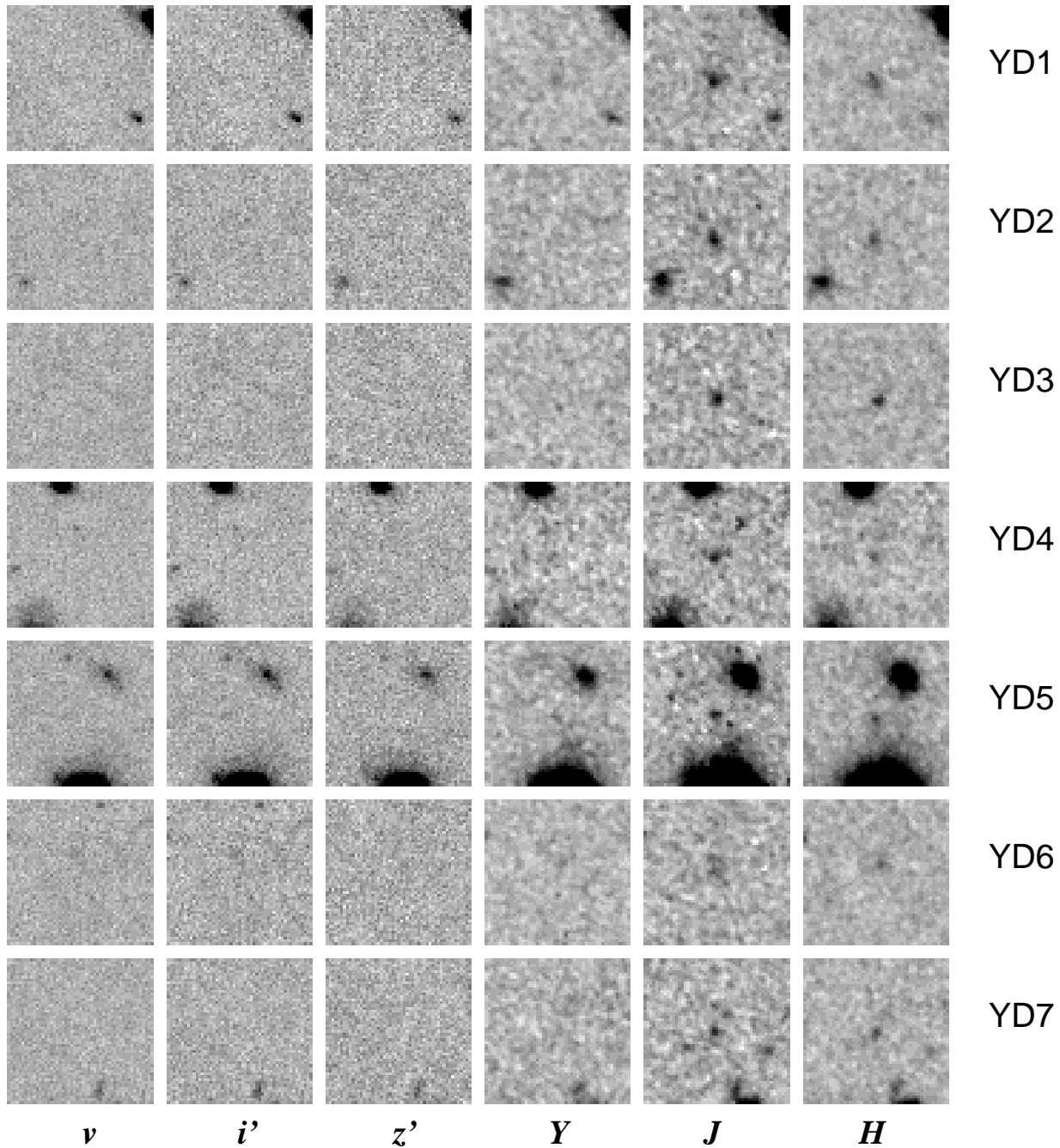}}
\caption{Postage stamp images of the $Y$-drops in the {\em Hubble} Ultra Deep Field.
The ACS $v$, $i'$, and $z'$ images are shown alongside the WFC3 $Y$, $J$ and $H$-bands 
for each object satisfying $(Y-J)_{AB}>1.0$, $J_{AB}<28.5$ and with no detection in the ACS images. Each image is 3\,arcsec across, with North up and East to the left.}
\label{fig:Ydrops}
\end{figure*}

The availability of deep longer-wavelength WFC3 images in addition to the $Y$-band enables us to push the Lyman break technique to higher redshifts still, searching for $z\approx 8$ objects where the Lyman-$\alpha$ decrement falls below the $J$-band (``$Y$-drops"), and $z\approx 10$ sources where the break falls between the $J$- and $H$-filters (``$J$-drops"). Again using SExtractor, but training in the $J$-images as our detection band, we uncover  7 objects with $(Y-J)_{AB}>1.0$, $J_{AB}<28.5$ and no ACS detection (Figure~\ref{fig:Ydrops} and Table~\ref{tab:ydrops}). These $Y$-drops
are potential $z\approx 8$ star-forming galaxies, and a discussion of the selection
and properties of these $z\approx 8$ candidates and the possible contamination fraction
will appear in a forthcoming paper (Lorenzoni et al.\ {\em in prep.})\footnote{{Note added in proof: an independent analysis of $Y$-drops in the WFC\,3-HUDF has been posted to the preprint server (Bouwens et al.\ submitted, arXiv:0909.1803); our sample contains all but one of the $Y$-drops presented in Bouwens et al., with UDFy-43086276 being fainter than our $J_{AB }<28.5$ cut. As with our analysis, Bouwens et al.\ find no robust $J$-drop candidate at $z\sim 10$.}}. Using $H$ as our detection band, we find no robust $(J-H_{AB})>1.0$ objects down to $H_{AB}<28.5$ which are undetected by ACS. The absence of robust $J$-drops hints at further evolution of the UV luminosity function at $z\gg7$.
  
\section{Conclusions}
\label{sec:concs}

We have searched for star-forming galaxies at $z\approx 7$ by applying the Lyman-break technique
to newly-released $1.1\,\mu$m $Y$-band images from WFC3 on the {\em Hubble} Space Telescope.
By comparing these images of the {\em Hubble} Ultra Deep Field with the ACS $z'$-band ($0.85\,\mu$m) images, we select objects with red colours, $(z'-Y)_{AB}>1.3$, consistent with the Lyman-$\alpha$ forest absorption at $z\approx 6.7-8.8$. We identify 12 of these $z'$-drops down to a limiting magnitude $Y_{AB}<28.5$ (equivalent to a star formation rate of $1.3\,M_{\odot}\,{\mathrm yr}^{-1}$ at $z=7.1$), all of which are undetected in the other ACS filters, consistent with the Lyman limit break at $z\approx 7$. We also analyse the new $1.25\,\mu$m $J$-band and 1.6\,$\mu$m $H$-band WFC3 images to measure the near-infrared colours of our $z'$-drops, and eliminate low mass Galactic stars which typically have redder colours than $z\approx 7$ galaxies. One of our $z'$-drops is a probably T-dwarf star. Our brightest $z'$-drop is not present in the NICMOS $J$-band image of the same field taken 5 years before, despite being well above the flux limit of this less sensitive image. This is a possible transient object, and we exclude this and the probable Galactic star. 

From the 10 remaining $z\approx 7$ candidates we determine a lower limit on the star formation rate density of $0.0017\,M_{\odot}\,{\mathrm yr}^{-1}\,{\mathrm Mpc}^{-3}$ for a Salpeter initial mass function, which rises to $0.0025-0.0034\,M_{\odot}\,{\mathrm yr}^{-1}\,{\mathrm Mpc}^{-3}$ after correction for luminosity bias. The star formation rate density is a factor of $\approx 10$ less than that of Lyman-break galaxies at $z=3-4$, and is about half the value at $z\approx 6$. This star formation rate density would produce insufficient Lyman continuum photons to reionize the Universe unless the escape fraction of these photons is extremely high ($f_{\mathrm esc}>0.5$) and the clumping factor of the Universe is low. Even then, we need to invoke a large contribution from galaxies below our detection limit (i.e., a steep faint end slope). The apparent shortfall in ionizing photons might be alleviated if stellar populations at  high redshift are low metallicity or have a top-heavy initial mass function. 

We have also inspected the $YJH$ images for the presence of Lyman-break galaxies at higher redshifts. We identify 7 objects with $(Y-J)_{AB}>1.0$ down to $J_{AB}<28.5$; these $Y$-drops are candidate $z\approx 8$ star-forming galaxies. A similar selection based on the $H$-band did not find any $J$-drops with $(J-H)_{AB}>1.0$ and $H_{AB}<28.5$ -- all such objects were detected in the ACS images (below the Lyman-limit at $z\sim 10$), and ruling out a high-redshift interpretation.

\subsection*{Acknowledgements}

We thank the anonymous referee for a careful reading of this manuscript and many helpful suggestions.
We are indebted to Elizabeth Stanway for her very significant input
in developing several of the tools used in this analysis. We thank Richard McMahon, Jim Dunlop, Ross McLure, Masami Ouchi, Bahram Mobasher and Michele Cirasuolo  for many useful discussions about Lyman break galaxies at high redshift. 
Based on observations made with the
NASA/ESA Hubble Space Telescope, obtained from the Data Archive at the
Space Telescope Science Institute, which is operated by the Association
of Universities for Research in Astronomy, Inc., under NASA contract NAS
5-26555. These observations are associated with programmes \#GO-11563, \#GO/DD-9978 and \#GO-9803.
SW and DS acknowledge funding from the U.K.\ Science and Technology Facilities Council. SH acknowledges a University of Hertfordshire studentship, and MJJ acknowledges a Research Councils UK Fellowship. SL is supported by the Marie Curie Initial Training Network ELIXIR of the European Commission under contract PITN-GA-2008-214227.

\bsp

\end{document}